\documentstyle[preprint,aps,floats,epsfig]{revtex}

\tightenlines

\def\appendix{\par
 \setcounter{section}{0}
 \setcounter{subsection}{0}
 \def\thesection{Appendix \Alph{section}}
 \def\thesubsection{\Alph{section}.\arabic{subsection}}
 \def\theequation{\Alph{section}.\arabic{equation}}
 \setcounter{equation}{0}}

\begin{document}

\renewcommand{\thefootnote}{\fnsymbol{footnote}}

\begin{flushright}
(version of June 13, 2000; to appear in Phys.~Lett.~B)
\end{flushright}

\centerline{{\large \bf 
Achieving renormalization--scale-- and scheme--independence}}
\centerline{{\large \bf 
in Pad\'e--related resummation in QCD}} 

\vspace{1.cm}

\centerline{G.~Cveti\v c}

\centerline{{\it Asia Pacific Center for Theoretical Physics}} 
\centerline{{\it 207-43 Cheongryangri-dong, Dongdaemun-gu, 
Seoul 130-012, Korea}} 
\centerline{{\it e-mail: cvetic@apctp.org}}

\renewcommand{\thefootnote}{\arabic{footnote}}

\begin{abstract}

Previously developed Pad\'e--related method of resummation
for QCD observables, which achieves exact 
renormalization--scale--invariance, is extended so that the 
scheme--invariance is obtained as well. The dependence on the 
leading scheme parameter $c_2$ is eliminated by a variant of the 
method of the principle of minimal sensitivity. The subleading
parameter $c_3$ in the approximant is then fixed in such
a way that the correct known location of the leading infrared
renormalon pole is reproduced. Thus, $\beta$--functions which
go beyond the last perturbatively calculated order in the
observable are used. The $\beta$--functions in the approximant 
are quasianalytically continued by Pad\'e approximants. 
Two aspects of nonperturbative physics are accounted 
for in the presented resummation: a mechanism of quasianalytic 
continuation from the weak-- into the strong--coupling regime, 
and the (approximant--specific) contribution of the leading 
infrared renormalon. The case of the Bjorken polarized sum rule is
considered as a specific example of how the method works.\\
PACS number(s): 11.10.Hi, 11.80.Fv, 12.38.Bx, 12.38.Cy

\end{abstract}

\setcounter{equation}{0}

\section{Introduction}

In QCD, as a result of extensive perturbative calculations, 
some observables are now known to the 
next--to--next--to--leading order (NNLO, $\sim$$a^3$) 
in the power expansion in the
strong coupling parameter $a\!\equiv\!{\alpha}_s/\pi$.
Knowing such truncated perturbation series (TPS),
the question of their resummation is gaining importance,
especially if the typical process energies associated with
the observable are low and thus the relevant coupling
parameter is large. In such cases, it is to be expected
that additional perturbative and nonperturbative effects, 
not explicitly contained in the TPS, will be numerically important. 
Many methods of resummation, based on the available TPS,
try to incorporate such effects.
Some of these methods eliminate the dependence
on the renormalization scale (RScl) 
and scheme (RSch) by fixing them in the TPS itself
in a judicious way -- these methods could be regarded
as renormalization--group--improved methods of resummation:
BLM--fixing motivated by the large--$n_f$ 
considerations \cite{BLM}, Stevenson's principle of minimal 
sensitivity (PMS) \cite{PMS}, Grunberg's effective charge method (ECH) 
\cite{ECH} (cf. Ref.~\cite{Gupta} for a related method).
Some of the recent works on resummations are
the method of ``commensurate scale relations'' and
related approaches \cite{csr}, 
a method using an analytic form
of the coupling parameter \cite{AQCD}, 
ECH--related approaches \cite{ECHrel}, 
and an expansion in the two--loop coupling parameter \cite{Kourashev}.
In the past few years, Pad\'e approximants (PA's) were also shown
to be a rather successful method of resummation \cite{Pade}, 
especially since the resummed results show
in general weakened RScl-- and RSch--dependence.
The diagonal Pad\'e approximants (dPA's) are particularly
well motivated for observables since they are 
RScl--independent in the approximation of the one--loop 
evolution of the coupling ${\alpha}_s(Q^2)$ \cite{Gardi}.
In addition, PA's go in their form beyond the polynomial form 
of the TPS on which they are based, thus contain a strong mechanism
of quasianalytic continuation from the weak-- into the
strong--coupling regime, and can consequently incorporate 
some of the nonperturbative effects into the resummed result.

Recently, an extension of the
method of dPA's was presented \cite{GC1}
which leads to the exact perturbative
RScl--independence of the resummed result. 
We then extended the method so that it is applicable also
to NNLO TPS's \cite{CK}, and suggested there
a way to fix the RSch by applying the principle
of minimal sensitivity (PMS).
The way suggested in \cite{CK} does not work properly in
practice since no minimum of the PMS equation
$\partial A/\partial c_2\!=\!0$ [Eq. (40) there] exists.
The dependence of our approximants on the RSch--parameters 
$c_2$ and $c_3$ of the original TPS
is a significant problem when the approximants are
applied to the low--energy
observables like the Bjorken polarized sum rule
(BjPSR) at low $Q_{\rm photon}\!\approx\!\sqrt{3}$ GeV \cite{GC2}.

This problem is addressed in the present paper.
For the case of NNLO TPS, 
an extended version ${\cal A}$ of our approximant 
is constructed where the dependence on the 
leading RSch--parameter $c_2$ is eliminated by a variant of the PMS. 
Subsequently, the sub--leading RSch--parameter $c_3$ is adjusted 
so that the approximant reproduces the correct location of the
leading infrared renormalon pole. The latter procedure
is carried out in the concrete example of the BjPSR. 
The same method of $c_3$--fixing is then applied to
Grunberg's ECH and Stevenson's TPS--PMS
approximants. 
Hence, in the approximants we use $\beta$--functions which
go beyond the last perturbatively calculated order 
in the observable (NNLO).
Further, a PA--type of quasianalytic continuation 
for the $\beta$--functions is used in all these approximants. 
The resulting predictions for  
${\alpha}_s^{{\overline {\rm MS}}}(M_Z^2)$ from the BjPSR
are presented, along with those when PA's are applied to the BjPSR,
and compared with the world average. Differences 
between between our approximant and the other methods 
(ECH, TPS--PMS; PA's) are pointed out. 

\section{Construction of $c_2$--independent approximants}

Consider a QCD observable $S$ 
with negligible mass effects and known NNLO TPS
\begin{eqnarray}
S_{[2]} &=& a_0 ( 1 + r_1 a_0 + r_2 a_0^2) \ ,
\label{S2TPS}
\\
{\rm with:} \quad 
a_0&\equiv& a(\ln Q_0^2;c_2^{(0)},c_3^{(0)},\cdots) \ , \quad
r_1 = r_1(\ln Q_0^2) \ , \quad
r_2 = r_2(\ln Q_0^2;c_2^{(0)}) \ .
\label{a0def}
\end{eqnarray}
We denoted $a\!\equiv\!\alpha_s/\pi$; 
$Q_0$ is the Euclidean RScl;
$c_j^{(0)}$ ($j\!\geq\!2$) are the RSch--parameters used in the TPS. 
The coupling parameter $a(\ln Q^2; c_2^{(0)}, \ldots)$
in this RSch evolves according to the renormalization 
group equation (RGE) $\partial a/ {\partial \ln Q^2}\!=\!\beta
(a; c_2^{(0)}, c_3^{(0)}, \ldots )$. Here,
the $\beta$--function has the power expansion
$\beta(a)\!=\!- \beta_0 a^2 
( 1\!+\!c_1 a\!+\!c_2^{(0)} a^2\!+\!c_3^{(0)} a^3\!+ \cdots )$,
and $\beta_0$ and $c_1$ are RScl-- and RSch--invariant. 
This RGE can be integrated (see Appendix of \cite{PMS})
\begin{equation}
{\beta}_0 \ln \left( \frac{Q_0^2}{{\tilde \Lambda}^2} \right) =
\frac{1}{a_0} + c_1 \ln 
\left( \frac{ c_1 a_0 }{ 1\!+\!c_1 a_0 } \right) +
\int_0^{a_0}\!dx \left[ \frac{1}{x^2(1\!+\!c_1 x)} +
\frac{ {\beta}_0 }{ {\beta}(x; c_2^{(0)}, c_3^{(0)} \ldots) } \right] \ ,
\label{Stevenson}
\end{equation}
where $a_0\!\equiv\!a(\ln Q^2_0; c_2^{(0)}, c_3^{(0)}, \ldots)$
and  ${\tilde \Lambda}$ is a universal scale ($\sim$$0.1$ GeV).
When subtracting (\ref{Stevenson}) from the analogous equation
for $a\!\equiv\!a(\ln Q^2; c_2, c_3, \ldots)$, 
an equation is obtained which relates $a$ with $a_0$,
i.e., determines $a$ in terms of $a_0$. This equation
then determines also the expansion of $a$ in powers of $a_0$.
{}From now on, we fix the ``$\Lambda$--convention'' to
${\Lambda}\!=\!{\tilde {\Lambda}}$.

We make the following ansatz for our approximant,
motivated by the RScl--invariant (but not RSch--invariant)
approximant of Refs.~\cite{CK,GC2}:
\begin{equation}
\sqrt{ {\cal A}_{S^2} } =
\left\{ {\tilde \alpha} \left[ 
a( \ln Q_1^2; c_2^{(1)}, c_3^{(1)}, \ldots )
- a( \ln Q_2^2; c_2^{(2)}, c_3^{(2)}, \ldots ) \right] 
\right\}^{1/2} 
\left( = S_{[2]} + {\cal O}(a_0^4) \right) \ ,
\label{Aans}
\end{equation}
where we regard now the parameters $c_2^{(j)}, c_3^{(j)}, \ldots$ 
($j\!=\!1,2$) as fixed numbers, and $c_2^{(1)}\!\not=\!c_2^{(2)}$.
Five parameters in
the approximant (${\tilde \alpha}$, $Q_1^2$, $Q_2^2$,
$c_2^{(1)}$, $c_2^{(2)}$) can be fixed by applying five
conditions to the approximant. Three conditions are obtained
from the so called minimal requirement: When we expand the
approximant back in powers of $a_0$, the first three
coefficients of the original TPS (\ref{S2TPS}) have to
be reproduced. The additional two conditions are obtained by
a variant of the PMS
\begin{equation}
( \partial {\cal A}_{S^2} /
\partial c_2^{(1)} ) {\big |}_{c_2^{(2)}}
\sim a_0^6 \sim
( \partial {\cal A}_{S^2} /
\partial c_2^{(2)} ) {\big |}_{c_2^{(1)}} \ .
\label{PMS}
\end{equation}
This allows us to fix $c_2^{(j)}$'s. 
If we took in (\ref{Aans}) $c_2^{(1)}\!=\!c_2^{(2)}$
($\equiv\!c_2$), and $c_k^{(1)}\!=\!c_k^{(2)}$ ($k\geq 3$),
i.e., the approximants of \cite{CK,GC2},
we would obtain 
$\partial {\cal A}_{S^2}/\partial c_2\!=\!-10 
c_1 a_0^5\!+\!{\cal O}(a_0^6) \not\sim a_0^6$,
i.e., the PMS condition would not be satisfied.
This is the main reason why we take 
two different (leading) parameters $c_2^{(j)}$ in the
two $a$'s in (\ref{Aans}). Further, since the two energy
scales in (\ref{Aans}), and in the approximants
of \cite{CK,GC2}, are $Q_1^2\!\not=\!Q_2^2$,
it does not appear unnatural to have 
$c_2^{(1)}\!\not=\!c_2^{(2)}$.
But the (subleading) parameters 
$c_3^{(j)}$ cannot be fixed by such an approach since
\begin{equation}
( \partial {\cal A}_{S^2} /
\partial c_3^{(s)} ) {\big |}_{{\delta}{c_3}}
= 2 a_0^5 + {\cal O}(a_0^6) \qquad 
{\rm where:} \
c_3^{(s)}\!\equiv\!(c_3^{(1)}\!+\!c_3^{(2)})/2, \
\delta c_3\!\equiv\!(c_3^{(1)}\!-\!c_3^{(2)}) \ .
\label{dc3}
\end{equation}
The same problem arises in the NNLO polynomial
approximants ECH and TPS--PMS where 
$\partial {\cal A_S}/\partial c_3\!=\!(1/2) a_0^4\!+\!{\cal O}(a_0^5)$
[$\Rightarrow \partial ({\cal A_S})^2/
\partial c_3\!=\!a_0^5\!+\!{\cal O}(a_0^6)$].
We will take, for simplicity, 
$c_3^{(1)}\!=\!c_3^{(2)}\!\equiv\!c_3$,
and the value of $c_3$ will be fixed later.

Conditions (\ref{PMS}) then depend also
on $\delta c_4\!\equiv\!(c_4^{(1)}\!-\!c_4^{(2)})$
which we set equal to zero to avoid further (presumably unnecessary)
complications. Then the set of the five equations determining 
${\tilde \alpha}$, $Q_j^2$ and $c_2^{(j)}$ ($j\!=\!1,2$) reads
\begin{eqnarray}
\lefteqn{
\!\!\!\!\!\!\!\!\!\!\!\!\!\!\!\!\!\!\!\!\!\!\!
y_{-}^4 -  y_{-}^2 z_0^2(c_2^{(s)}) +
y_{-} \frac{5}{4} c_1 \delta c_2 
- \frac{3}{16} (\delta c_2)^2 =  0 \ , }
\label{eq1}
\\
\lefteqn{
\!\!\!\!\!\!\!\!\!\!\!\!\!\!\!\!\!\!\!\!\!\!\!
{\bigg \{}
 27 (\delta c_2)^3 - 157 c_1 (\delta c_2)^2 y_{-}
- 8 \delta c_2 y_{-}^2 \left[ - 27 c_1^2\!+\!12 c_2^{(s)}
\!+\!34 y_{-}^2\!-\!8 z_0^2(c_2^{(s)}) \right]
}
\nonumber\\
&& + 48 c_1 y_{-}^3  
\left[ 13 y_{-}^2\!-\!3 z_0^2(c_2^{(s)}) \right] {\bigg \}} 
\left[
5 c_1 \delta c_2\!+\!16 y_{-}^3\!-\!8 z_0^2(c_2^{(s)}) y_{-} )
\right]^{-1} = 0 ,
\label{PMSeq1}
\\
\lefteqn{
\!\!\!\!\!\!\!\!\!\!\!\!\!\!\!\!\!\!\!\!\!\!\!
{\bigg \{}
27 (\delta c_2)^4 - 315 c_1 (\delta c_2)^3 y_{-}
+ 64 z_0^4(c_2^{(s)}) y_{-}^2 \left[ 7 c_1^2\!-\!2 c_2^{(s)}
\!+\!3 z_0^2(c_2^{(s)}) \right] 
-4 \delta c_2 (20 c_1 y_{-}\!-\!3 \delta c_2) 
}
\nonumber\\
&& \times
\left[ - 2 c_2^{(s)} y_{-}^2
\!-\!2 c_2^{(s)} z_0^2(c_2^{(s)})\!+\!12 z_0^2(c_2^{(s)}) y_{-}^2
\!+\!3 z_0^4(c_2^{(s)})\!+\!7 c_1^2 
\left( y_{-}^2 \!+\!z_0^2(c_2^{(s)}) \right)
\right]
\nonumber\\
&& + 36 (\delta c_2)^2 y_{-}^2 \left[ 
z_0^2(c_2^{(s)})\!+\!25 c_1^2 \right] 
{\bigg \}}
\left[
5 c_1 \delta c_2\!+\!16 y_{-}^3\!-\!8 z_0^2(c_2^{(s)}) y_{-}
\right]^{-1} = 0 \ ,
\label{PMSeq2}
\\
\lefteqn{
\!\!\!\!\!\!\!\!\!\!\!\!\!\!\!\!\!\!\!\!\!\!\!
- r_1 + \frac{1}{2} c_1 - \frac{1}{4} \frac{\delta c_2}{y_{-}}
 =  y_{+} \ , \qquad
{\tilde \alpha} =  - \frac{1}{2 y_{-}} \ , }
\label{eq1b}
\end{eqnarray}
where we use notations
\begin{eqnarray}
y_{\pm} &\equiv& \frac{1}{2} \beta_0 \left[ \ln \frac{Q_1^2}{Q_0^2}
\pm \ln \frac{Q_2^2}{Q_0^2} \right] \ , \quad
\delta c_2  \equiv  c_2^{(1)} - c_2^{(2)} \ , \quad
c_2^{(s)} \equiv \frac{1}{2}(c_2^{(1)} + c_2^{(2)}) \ ,
\label{notf}
\\
z_0^2 &\equiv& \left( 2 \rho_2\!+\!\frac{7}{4} c_1^2 \right)
- 3 c_2^{(s)} \equiv z_0^2(c_2^{(s)}) \ , \qquad 
\rho_2 = r_2 - r_1^2 - c_1 r_1 + c_2^{(0)} \ .
\label{z0}
\end{eqnarray}
Here, $\rho_2$ is an RScl-- and RSch--invariant, and therefore
it is straightforward to see that the solutions of the
system (\ref{eq1})--(\ref{eq1b}) for $Q_j^2$ and
$c_2^{(j)}$ ($j\!=\!1,2$) and for $\tilde \alpha$ are
independent of the original choice of the RScl ($Q_0^2$)
and of the RSch ($c_2^{(0)}, c_3^{(0)}, \ldots$).
Eqs.~(\ref{PMSeq1})--(\ref{PMSeq2}) originate from
PMS conditions (\ref{PMS}), and the other three
identities from the minimal condition.
In particular, the latter three identities 
(\ref{eq1}) and (\ref{eq1b}) show that
${\tilde {\alpha}}$ and the scales $Q_1^2$ and $Q_2^2$
are $Q_0^2$--independent irrespective of whether
$\delta c_2\!\not=\!0$ or $\delta c_2\!=\!0$.

The coupled system of three equations
(\ref{eq1})--(\ref{PMSeq2}) for the three unknowns
$c_2^{(j)}$ ($j\!=\!1,2$) and 
$y_{-}\!\equiv\!{\beta}_0 \ln(Q_1/Q_2)$
can be solved numerically. The solutions which
give $|{\tilde \alpha}|\!\ll\!1$
or $|{\tilde \alpha}|\!\gg\!1$ must be
discarded because they would cause numerical instabilities
in the approximant, and they would not make sense
physically either -- one of the scales $Q_1$, $Q_2$
would be orders of magnitude different from the other.
There are apparently two possibilities:
1.) $y_{-}$, $c_2^{(s)}$ and $\delta c_2$ are all real numbers;
2.) $c_2^{(s)}$ is real,
$y_{-}$ and $\delta c_2$ are imaginary numbers.
In both cases, the approximant itself would be real,
as it shoud be.

If there are several solutions which
give different values for the approximant, we should
choose (again within the PMS--logic) among them the
solution with the smallest curvature with respect
to $c_2^{(1)}$ and $c_2^{(2)}$.

\section{Application to the Bjorken polarized sum rule; $c_3$-fixing}

The Bjorken polarized sum rule (BjPSR)
involves the isotriplet combination of the 
first moments over $x_{\rm Bj}$ of proton and
neutron polarized structure functions
\begin{equation}
\int_0^1 d x_{\rm Bj} \left[ g_1^{(p)} (x_{\rm Bj}; Q^2_{\rm ph})
- g_1^{(n)} (x_{\rm Bj}; Q^2_{\rm ph}) \right] =
 \frac{1}{6} |g_A| \left[ 1 - S(Q^2_{\rm ph}) \right] \ ,
\label{BjPSR1}
\end{equation}
where $p^2\!=\!-Q^2_{\rm ph}$$<0$ is $\gamma^{\ast}$ 
momentum transfer. At $Q^2_{\rm ph}\!=\!3 {\rm GeV}^2$ 
where three quarks are assumed active ($n_f\!=\!3$), 
and if taking ${\overline {\rm MS}}$ RSch and
RScl $Q_0^2\!=\!Q^2_{\rm ph}$, we have \cite{GLZN,LV}:
\begin{eqnarray}
S_{[2]}(Q^2_{\rm ph}; Q^2_0 = Q^2_{\rm ph}; 
c_2^{{\overline {\rm MS}}}, c_3^{{\overline {\rm MS}}})
&=& a_0 ( 1 + 3.583 a_0 + 20.215 a_0^2) \ ,
\label{TPSBj}
\\
{\rm with:} \quad a_0 = 
a(\ln Q_0^2; c_2^{{\overline {\rm MS}}}, 
c_3^{{\overline {\rm MS}}}, \ldots) \ , \quad 
n_f&=&3 \ , \ c_2^{{\overline {\rm MS}}} = 4.471, \ 
c_3^{{\overline {\rm MS}}} = 20.99 \ .
\label{TPSBjnot}
\end{eqnarray}
Solving numerically the system of equations
(\ref{eq1})--(\ref{eq1b}), we obtain one solution only
\begin{equation}
c_2^{(1)} = 1.465 \ , \ c_2^{(2)} = 5.137 \ , \
Q_1 = 0.594 \ {\rm GeV} \ , Q_2 = 1.164 \ {\rm GeV} \quad
(\Rightarrow {\tilde \alpha} = 0.3301) \ .
\label{PMSsol}
\end{equation}
This solution is independent of the choice of RScl and RSch.
For the time being, we will set the higher parameters
$c_k^{(j)}\!=\!0$ ($k \geq 4$, $j\!=\!1,2$).
Now our approximant depends only on the still free parameter
$c_3$. This dependence is numerically significant.
For a typical value $a_0\!=\!0.09$ [$\Rightarrow$
$\alpha_s^{{\overline {\rm MS}}}(3 {\rm GeV}^2)\!\approx\!0.283$,
$\alpha_s^{{\overline {\rm MS}}}(M_Z^2)\!\approx\!0.113$],
the approximant (\ref{Aans}) gives $0.1523$ and $0.1632$
when $c_3\!=\!0, c_3^{{\overline {\rm MS}}}$, respectively, 
i.e. a difference of $7.2 \%$.
In the case of the ECH and TPS--PMS approximants for the BjPSR,
the respective differences are $3.8 \%$ and $4.0 \%$.

Parameter $c_3$ characterizes the ${\rm N}^3 {\rm LO}$
term in the corresponding $\beta$--functions
[cf.~Eq.~(\ref{Stevenson})], and information on its value cannot
be obtained from the NNLO TPS [cf.~Eq.~(\ref{dc3})].
Therefore, to fix $c_3$, we should incorporate into
the approximants a known piece of (nonperturbative) 
information beyond the NNLO TPS (\ref{TPSBj}). 
Natural candidates for this are the known locations of the 
poles \cite{BK,TM} of the leading infrared renormalon 
(${\rm IR}_1$: $z_{\rm pole}\!=\!1/{\beta}_0$)
or ultraviolet renormalon 
(${\rm UV}_1$: $z_{\rm pole}\!=\!-1/{\beta}_0$),
i.e., the poles of the Borel transform $B_S(z)$
of $S$ closest to the origin. 
Large--$\beta_0$ evaluations \cite{TM}, based on the formulas
of \cite{BK} and using simple Borel transforms in a
variant of the V--scheme (RScl $Q_0\!=\!Q_{\rm ph} \exp(-5/6)$
and one--loop--evolved $a$), suggest that the ${\rm UV}_1$
contributions to the BjPSR at 
$Q^2_{\rm ph}\!=\!2$-$3 \ {\rm GeV}^2$
are suppressed in comparison to the ${\rm IR}_1$
contributions by a factor 3--4 (cf.~their Fig.~2).

Therefore, we will fix $c_3$ in the three approximants
by incorporating in them
the information on the location of the
${\rm IR}_1$ pole $z_{\rm pole}\!=\!1/{\beta}_0$ ($=\!4/9$).
For that, we employ RScl-- and RSch--invariant Borel transforms.
Simple Borel transforms are not RScl/RSch--invariant,
the use of their TPS's leads to RScl/RSch--dependent
$c_3$--fixing, which we want to avoid.
We use a variant of the invariant Borel 
transform $B(z)$ introduced by Grunberg \cite{BTinv}, who in turn
introduced it on the basis of the modified Borel transform
of Ref.~\cite{BTmod}
\begin{equation}
S(Q^2_{\rm ph}) = \int_0^{\infty} dz 
\exp \left[ - {\rho}_1 (Q^2_{\rm ph}) z \right] B_{S}(z) \ ,
\label{BT1}
\end{equation}
where ${\rho}_1$ is the first RScl/RSch--invariant \cite{PMS}
of $S$:
${\rho}_1\!=\!-r_1\!+\!{\beta}_0 \ln 
(Q_0^2/ {\tilde \Lambda}^2)\!=\!{\beta}_0 \ln 
(Q^2_{\rm ph}/{\overline \Lambda}^2)$.
Here, ${\tilde \Lambda}$ is the universal scale
of Eq.~(\ref{Stevenson}),
and ${\overline \Lambda}$ a scale which depends
on the choice of $S$ but is RScl/RSch--invariant and 
$Q_{\rm ph}$--independent. 
The ${\rho}_1(Q^2_{\rm ph})$ is, 
up to an additive constant (the latter not affecting
the poles of $B_S$), equal to 
$1/a^{\rm (1-loop)}(Q^2_{\rm ph})$.
Thus, $B_S(z)$ of (\ref{BT1}) reduces to the simple
Borel transform, up to a factor $\exp(c z)$, once higher
than one--loop effects are ignored.
The coefficients of the
power expansion of $B_S(z)$ of (\ref{BT1}) are
RScl/RSch--invariant, in contrast to the case
of the simple Borel transform.
These invariant coefficients can be related with
coefficients $r_n$ of $S$
most easily in a specific RSch $c_k\!=\!c_1^k$ ($k \geq 2$)
\begin{equation}
B_S(z) = (c_1 z)^{c_1 z} \exp( - r_1 z)
\sum_0^{\infty} 
\frac{ ( {\tilde r}_n\!-\!c_1 {\tilde r}_{n-1} ) }
{ {\Gamma}(n\!+\!1\!+\!c_1 z) } z^n 
\equiv (c_1 z)^{c_1 z} {\overline B}_S(z) \ .
\label{BT3}
\end{equation}
Here, ${\tilde r}_n$ is the coefficient at ${\tilde a}^{n+1}$
in the expansion of $S$ in powers of
${\tilde a}\!\equiv\!a(\ln Q_0^2;c_1^2,c_1^3,...)$;
by definition
${\tilde r}_{-1}\!=\!0$, ${\tilde r}_{0}\!=\!1$.
Thus, the expansion of the approximant $\sqrt{{\cal A}_{S^2}}(c_3)$
in powers of ${\tilde a}$ leads to the expansion
of the (reduced) Borel transform 
${\overline B}_{{\sqrt {\cal A}}}(z)$ in powers of $z$.
The coefficients starting at $z^3$ are predictions of
the approximant and $c_3$--dependent:
${\overline B}_{{\sqrt {\cal A}}}(z)\!=\!1\!+\!{\bar b}_1 
z\!+\!{\bar b}_2 z^2\!+\!{\bar b}_3 z^3\!+\cdots$, with 
${\bar b}_1\!\approx\!-0.7516$,
${\bar b}_2\!\approx\!0.4209$,
${\bar b}_3\!\approx\!(-2.664\!+\!0.1667 c_3)$, etc.
Terms with high powers of $z$ are not reliable, because the
approximant is based on an NNLO TPS $S_{[2]}$
with only two terms beyond the leading order.
We then employ Pad\'e approximants (PA's) of
power expansion of ${\overline B}_{{\sqrt {\cal A}}}$, 
since they are efficient in determining the pole 
structure of ${\overline B}_{{\sqrt {\cal A}}}$.
We performed the expansion of $\sqrt{\cal A}(c_3)$ 
up to $\sim$${\tilde a}^7$, obtaining
the expansion of ${\overline B}_{{\sqrt {\cal A}}}(z)$
up to $\sim$$z^6$. This allowed us to construct 
${\rm PA}_{\overline B}$'s of as high order
as $[3/3]$ or $[4/2]$. The value of $c_3$ in
${\rm PA}_{\overline B}$ was then adjusted to achieve 
$z_{\rm pole}\!=\!1/{\beta}_0$ ($=\!4/9$). The resulting
values of $c_3$ are presented in the second column (${\rm TPS}_{\beta}$)
of Table \ref{tabl1}. 
\begin{table}[ht]
\par
\begin{center}
\begin{tabular}{l || c c | c c | c c}
${\rm PA}_{\overline B}$  & 
$c_3$ ($\sqrt{{\cal A}_{S^2}}$): ${\rm TPS}_{\beta}$ & ${\rm PA}_{\beta}$ &
$c_3$ (ECH): ${\rm TPS}_{\beta}$                     & ${\rm PA}_{\beta}$ &
$c_3$ (TPS--PMS): ${\rm TPS}_{\beta}$                & ${\rm PA}_{\beta}$ \\
\hline \hline
$[2/1]$ & 21.7 & 21.7 & 35.1 & 35.1 & 35.1 & 35.1 \\
\hline
$[3/1]$ & 13.7 & 15.7 & 19.5 & 22.9 & 19.0 & 21.5 \\
\hline
$[4/1]$ & 11.1 & 15.8 & 14.4 & 20.8 & 13.1 & 18.7 \\
\hline
$[5/1]$ & 9.3  & 16.9 & 11.2 & 19.6 &  8.8 & 17.3 \\
\hline\hline
$[1/2]$ & 12.8 & 12.8 & 17.3 & 17.3 & 17.3 & 17.3 \\
\hline
$[2/2]$ & 12.4 & 14.9 & 16.9 & 20.4 & 16.2 & 19.4 \\
\hline
$[3/2]$ & $11.7\!\pm\!3.4 {\rm i}$ & 15.8 &
$15.8\!\pm\!6.4 {\rm i}$ & $20.7\!\pm\!2.8 {\rm i}$ &        
$15.4\!\pm\!7.4 {\rm i}$ & $17.3\!\pm\!3.6 {\rm i}$ \\
\hline
$[4/2]$ & $10.3\!\pm\!2.8 {\rm i}$ & 15.7 &
$12.9\!\pm\!5.1 {\rm i}$ & $20.4\!\pm\!1.8 {\rm i}$ &
$11.6\!\pm\!6.8 {\rm i}$ & $17.0\!\pm\!2.6 {\rm i}$ \\
\hline\hline
$[1/3]$ & 12.4 & 15.0 & 16.9 & 20.6 & 16.2 & 19.5 \\
\hline
$[2/3]$ & 12.9 & $15.1\!\pm\!1.2 {\rm i}$ &
17.4 &                               19.3 &
$18.3\!\pm\!0.8 {\rm i}$ &           18.5 \\
\hline
$[3/3]$ & 
$10.6\!\pm\!2.9 {\rm i}$ & $14.0\!\pm\!1.7 {\rm i}$ &
$13.6\!\pm\!5.5 {\rm i}$ & $20.2\!\pm\!2.0 {\rm i}$ &
$12.6\!\pm\!7.0 {\rm i}$ & $16.9\!\pm\!2.7 {\rm i}$
\end{tabular}
\end{center}
\caption {\footnotesize Predictions for $c_3$ in our, ECH and
TPS--PMS approximants, using PA's of the invariant
Borel transform ${\overline B}(z)$ of the approximants
and demanding that the ${\rm IR}_1$ pole be at
$z_{\rm pole}\!=\!1/{\beta}_0$ ($=4/9$). ``${\rm TPS}_{\beta}$''
denotes that the parameters
$c_k^{(j)}$ ($k \geq 4$, $j\!=\!1,2$) in 
$\sqrt{{\cal A}_{S^2}}$,
and $c_k$ ($k \geq 4$) in ECH and TPS--PMS, are set
equal to zero; ``${\rm PA}_{\beta}$'' denotes 
that the $\beta$--functions in the approximants are resummed as:
$[2/3]_{\beta}$ (RSch1);
$[2/4]_{\beta}$ (RSch2; $c_4^{(2)}\!=\!c_4^{(1)}$);
$[3/2]_{\beta}$ (ECH RSch, and TPS--PMS RSch).}
\label{tabl1}
\end{table}
We carried out the analogous 
$c_3$--fixing for the polynomial approximants
ECH and TPS--PMS\footnote{
The ECH approximant is 
${\cal A}_S^{({\rm ECH})}(c_3)\!=\! 
a(\ln Q^2_{\rm ECH}; {\rho}_2, c_3, \ldots)$;
the TPS--PMS approximant is
${\cal A}_S^{({\rm PMS})}(c_3)\!=\!
a_{\rm PMS}\!-\!{\rho}_2 a^3_{\rm PMS}/2$, with 
$a_{\rm PMS}(c_3)\!=\!a(\ln Q^2_{\rm ECH}; 3 {\rho}_2/2, c_3, \ldots)$,
$Q^2_{\rm ECH}\!=\!Q^2_0 \exp(-r_1/{\beta}_0)$.}
to the BjPSR, and $c_3$ predictions for them
are also included in
Table \ref{tabl1} (columns with ``${\rm TPS}_{\beta}$''). 
These entries in the Table suggest
the values $c_3 \approx 12.5, 17, 16$ for $\sqrt{{\cal A}_{S^2}}$,
ECH, and TPS--PMS, respectively. The
predictions of ${\rm PA}_{\overline B}$'s of 
intermediate order 
($[3/1]$, $[4/1]$, $[2/2]$, $[3/2]$, $[1/3]$, $[2/3]$) 
appear to give the most stable results. Predictions of the
higher order ${\rm PA}_{\overline B}$'s 
gradually lose predictability
(predicted $c_3$'s can even become complex) because
of the afore--mentioned overdetermination.
The lowest order ${\rm PA}_{\overline B}$'s 
are unreliable due to their too simple structure.

The possibility to adjust the ${\rm N}^3 {\rm LO}$ coefficient
$r_3$ of (\ref{TPSBj}) in a similar way,
was apparently first mentioned by the authors of Ref.~\cite{EGKS}.
They referred to PA's ($[2/1]$) of the simple Borel transform,
so their (PA--resummed) predictions would depend on
the choice of the RScl and RSch.
A systematic method to optimize the perturbative expansion
by including the information on the location of the 
${\rm IR}_1$ pole was suggested in Ref.~\cite{CapriniFischer}.

Up until now we have taken the higher order parameters
$c_k^{(j)}$ ($k \geq 4$, $j\!=\!1,2$) in our approximant
(and in the ECH and TPS--PMS) to be zero, thus truncating
the corresponding $\beta$--functions (${\rm TPS}_{\beta}$). 
However, since the considered observable
has low process energy $Q_{\rm ph} \approx 1.73$ GeV,
we expect the higher order terms $\propto\!c_k^{(j)} x^{k+2}$
($k \geq 4$, $x\!\equiv\!{\alpha}_s/\pi$) of the
$\beta$--function to contribute significantly to the determination 
(via evolution) of the relevant coupling parameters
of the approximants. This leads us immediately to the
question of quasianalytic continuation of the $\beta(x)$
functions from the small--$x$ into the large--$x$ regime.
We can choose again Pad\'e approximants (PA's) as a tool
of this quasianalytic continuation, keeping $c_3$ as the
only free parameter, and subsequently
determine $c_3$ in the afore--mentioned way. 

In $\sqrt{ {\cal A}_{S^2} }(c_3)$
there are $\beta$--functions characterized
by the RSch--parameters $(c_2^{(1)}, c_3,\ldots)$ (RSch1) and 
$(c_2^{(2)}, c_3,\ldots)$ (RSch2) and determining the evolution 
and values of 
$a_1\!\equiv\!a(\ln Q_1^2; c_2^{(1)}, c_3, \ldots)$ and
$a_2\!\equiv\!a(\ln Q_2^2; c_2^{(2)}, c_3, \ldots)$,
respectively. In the (NNLO) ECH and the TPS--PMS approximants,
the RSch--sets are $(\rho_2, c_3, \ldots)$
and $(3 \rho_2/2, c_3, \ldots)$, respectively.
For RSch1, ECH and TPS--PMS RSch, we have at first the
freedom to construct $[2/3]$, $[3/2]$, or $[4/1]$ 
${\rm PA}_{\beta}$'s. For RSch2, the additional condition 
$c_4^{(2)}\!=\!c_4^{(1)}$ ($\delta c_4\!=\!0$) has to be fulfilled. 
Since $c_4^{(1)}$ is a unique function of $c_3$ once a 
${\rm PA}_{\beta1}$ choice has been made for RSch1, 
we then have for ${\rm PA}_{\beta2}$ of RSch2 
the possibilities $[2/4], [3/3], [4/2], [5/1]$.
For each choice of ${\rm PA}_{\beta}$'s, we essentially
repeat the afore--mentioned procedure of determining the value
of $c_3$. We consider the best choice of ${\rm PA}_{\beta}$'s 
the one giving the most stable prediction of $c_3$
over various PA's $[M/N]_{\overline B}$ of the 
approximant's invariant Borel transform.
This turns out to be for 
$\sqrt{ {\cal A}_{S^2} }(c_3)$ the choice
($[2/3]_{\beta1}$, $[2/4]_{\beta2}$),
although ($[2/3]_{\beta1}$, $[5/1]_{\beta2}$)
give virtually the same and almost as stable $c_3$--predictions.
For the ECH and TPS--PMS the choice is $[3/2]_{\beta}$.
The predictions for $c_3$ are given in Table 
\ref{tabl1} (columns with ``${\rm PA}_{\beta}$''). 
Those from PA's $[M/N]_{\overline B}$
of intermediate order
are significantly more stable than the corresponding ones
with truncated $\beta$--functions (``${\rm TPS}_{\beta}$''). 
This is a numerical indication
that the PA--resummation of the $\beta$--functions 
improves the ability of the
approximants to discern nonperturbative effects in
the considered observable. The ``${\rm PA}_{\beta}$''--entries 
in Table \ref{tabl1}
give us approximate values $c_3\!=\!15.5, 20, 19$ for
our, the ECH, and the TPS--PMS approximant, respectively.

There is yet another argument in favor of the above
${\rm PA}_{\beta}$ choices. The chosen
$[2/3]_{\beta1}$ and $[2/4]_{\beta2}$ (or: $[5/1]_{\beta2}$)
have positive poles with mutually similar values: 
$x_{\rm pole}=0.334, 0.325$ (or: $0.291$), respectively.
The value of $x_{\rm pole}$ ($=\!{\alpha}_{\rm pole}/\pi$)
indicates a point where ``a strong and an 
asymptotically--free phase share a common infrared attractor''
\cite{Chishtieetal}. Thus, it is reasonable to expect that
only those RSch's whose $\beta(x)$--functions have about the
same value of $x_{\rm pole}$ are suitable for the use
in calculation of nonperturbative effects 
(on the other hand, in purely perturbative QCD, 
all RSch's are formally equivalent).
Hence, the mutual proximity of $x_{\rm pole}$'s of
RSch1 and RSch2 ${\rm PA}_{\beta}$'s is now
yet another indication that these ${\rm PA}_{\beta}$'s
are the reasonable ones. 
What happens if we choose for RSch1 and RSch2 
other ${\rm PA}_{\beta}$'s?
In such cases, we always end up with one of the following
situations: Either the two corresponding
positive $x_{\rm pole}$ values are far apart,
or both values are unphysically small, 
or one (positive) $x_{\rm pole}$ doesn't exist, 
or there are no predictions for $c_3$ (not even unstable),
or $x_{\rm pole}$ values are unstable under the
change of $c_3$ in the interesting region 
$c_3 \approx 12$--$16$. 
So, the choice $[2/3]_{\beta1}$ and 
$[2/4]_{\beta2}$ (or $[5/1]_{\beta2}$)
in our approximant is not just the choice giving the 
most stable $c_3$--predictions, it is also the only choice
giving mutually similar (and reasonable) values of $x_{\rm pole}$
of RSch1 and RSch2.
Further, the choice
$[3/2]_{\beta}$ for the ECH and TPS--PMS RSch's
gives us $x_{\rm pole}$ values similar
to the ones previously mentioned:
$x_{\rm pole}\!=\!0.263$ for ECH with $c_3\!=\!20$;
$x_{\rm pole}\!=\!0.327$ for TPS--PMS with $c_3\!=\!19$.
Even other choices of ${\rm PA}_{\beta}$
for the ECH and TPS--PMS RSch's ($[2/3]_{\beta}$, $[4/1]_{\beta}$),
which also give rather stable and very similar $c_3$--predictions,
give us $x_{\rm pole} \approx 0.27$--$0.41$.
Therefore, we see in all cases a clear correlation
between the stability of the $c_3$--predictions
on the one hand and $x_{\rm pole} \approx 0.3$--$0.4$
on the other hand.
Finally, $[2/3]_{\beta}$ is then the good
choice for ${\overline {\rm MS}}$ RSch
since it has $x_{\rm pole}\!=\!0.311$
($c_3^{{\overline {\rm MS}}}$, cf.~Eq.~(\ref{TPSBjnot}),
has been determined in Ref.~\cite{RVL}). 
The choices $[3/2]_{\beta}$ and $[4/1]_{\beta}$ 
for  ${\overline {\rm MS}}$ 
give $x_{\rm pole}\!=\!0.119, 0.213$, 
respectively, which are further away from $0.3$--$0.4$.

Now that all the hitherto unknown parameters in 
$\sqrt{{\cal A}}$ of (\ref{Aans}) and in the
ECH and TPS--PMS have been determined, we
use the approximants to predict the values of 
$\alpha_s^{{\overline {\rm MS}}}(3 {\rm GeV}^2)$
($=\!\pi a_0$) from the measured values of the BjPSR 
$S(Q^2_{\rm ph}\!=\!3 {\rm GeV}^2)$.
Experimental values at $Q_{\rm ph}\!=\!\sqrt{3}$ GeV
are given in \cite{ABFR} (their Table 4) and are based on
SLAC data
\begin{equation}
\frac{1}{6} |g_A| \left[ 1 - S(Q^2_{\rm ph}) \right] = 
0.177 \pm 0.018 \  \qquad 
\Rightarrow \
S(Q^2_{\rm ph}) = 0.155 \pm 0.086 \ .
\label{Bjdat}
\end{equation}
where the constant $|g_A|$ is known \cite{ABFR} from $\beta$--decay
measurements: $|g_A|\!=\!1.257 \ (\pm 0.2 \%)$.
The experimental uncertainties are high, 
mainly because of the effects of perturbative
evolution on the small--$x_{\rm Bj}$ extrapolation
of the polarized structure functions appearing in
the sum rule (\ref{BjPSR1}), as explained in Ref.~\cite{ABFR}. 
We vary $a_0$ in our, and any other, approximant
for the BjPSR $S$ in such a way that the values (\ref{Bjdat})
are reproduced. We then obtain the predictions
for ${\alpha}_s^{{\overline {\rm MS}}}(3 {\rm GeV}^2)$ 
given in Table \ref{tabl2}. 
Given are always three values for $\alpha_s$,
corresponding to the three values of $S$ (\ref{Bjdat}). 
The results are given for
our, the ECH and the TPS-PMS approximants,
all with the described $c_3$--fixing and with the afore--mentioned
PA--type resummation of the pertaining $\beta$--functions:
$[2/3]_{\beta}$ (RSch1), $[2/4]_{\beta}$ (RSch2),
$[3/2]_{\beta}$ (ECH), $[3/2]_{\beta}$ (TPS--PMS),
$[2/3]_{\beta}$ (${\overline {\rm MS}}$).
\begin{table}[ht]
\par
\begin{center}
\begin{tabular}{l c c}
approximant & $\alpha_s( 3 \ {\rm GeV}^2)$ & $\alpha_s(M_Z^2)$ \\
\hline \hline
$\sqrt{ {\cal A}_{S^2} }$ ($c_3\!=\!15.5$; ${\rm PA}_{\beta}$'s) & 
$0.2755^{+0.0342}_{-0.1068}$ &
$0.1120^{+0.0047}_{-0.0219}$ \\
\hline
ECH ($c_3\!=\!20.$; ${\rm PA}_{\beta}$'s) &  
$0.2770^{+0.0371}_{-0.1082}$ &
$0.1122^{+0.0051}_{-0.0221}$ \\
\hline
TPS--PMS ($c_3\!=\!19.$; ${\rm PA}_{\beta}$'s) & 
$0.2778^{+ {\rm ?}}_{-0.1090}$ &
$0.1123^{+ {\rm ?}}_{-0.0222}$ \\ 
\hline\hline
$\sqrt{ {\cal A}_{S^2} }$ ($c_3\!=\!12.5$; ${\rm TPS}_{\beta}$'s) & 
$0.2798^{+0.0487}_{-0.1109}$ &
$0.1126^{+0.0064}_{-0.0224}$ \\
\hline
ECH ($c_3\!=\!17.$; ${\rm TPS}_{\beta}$'s) &
$0.2801^{+0.0504}_{-0.1112}$ &
$0.1127^{+0.0066}_{-0.0225}$ \\
\hline 
TPS--PMS ($c_3\!=\!16.$; ${\rm TPS}_{\beta}$'s) & 
$0.2808^{+ {\rm ?}}_{-0.1119}$ &
$0.1128^{+ {\rm ?}}_{-0.0226}$ \\
\hline\hline
$\sqrt{ {\cal A}_{S^2} }$ ($c_3\!=\!0.$; ${\rm TPS}_{\beta}$'s) & 
$0.2853^{+0.0581}_{-0.1159}$ &
$0.1134^{+0.0073}_{-0.0231}$ \\
\hline
ECH ($c_3\!=\!0.$; ${\rm TPS}_{\beta}$'s) &
$0.2841^{+0.0573}_{-0.1148}$ &
$0.1133^{+0.0072}_{-0.0230}$ \\
\hline
TPS--PMS ($c_3\!=\!0.$; ${\rm TPS}_{\beta}$'s) & 
$0.2848^{+ {\rm ?}}_{-0.1155}$ &
$0.1134^{+ {\rm ?}}_{-0.0231}$ \\
\hline \hline
$[2/2]_S$ (${\rm N}^3{\rm LO}$, $r_3\!=\!128.05$)  & 
$0.2838^{+0.0595}_{-0.1147}$ &
$0.1132^{+0.0075}_{-0.0230}$ \\
\hline
$\sqrt{[2/2]_{S^2}}$ & 
$0.2832^{+0.0569}_{-0.1141}$ &
$0.1131^{+0.0071}_{-0.0229}$ \\
\hline
$[2/1]_S$ &
$0.2890^{+0.0671}_{-0.1194}$ &
$0.1140^{+0.0080}_{-0.0237}$ \\
\hline
$[1/2]_S$ &
$0.2930^{+0.0727}_{-0.1230}$ &
$0.1145^{+0.0085}_{-0.0240}$ \\
\hline
${\rm N}^3{\rm LO}$ TPS ($r_3\!=\!128.05$) & 
$0.2983^{+0.0855}_{-0.1281}$ &
$0.1152^{+0.0095}_{-0.0247}$ \\
\hline
NNLO TPS & 
$0.3127^{+0.1021}_{-0.1403}$ & 
$0.1171^{+0.0102}_{-0.0260}$
\end{tabular}
\end{center}
\caption {\footnotesize Predictions for 
${\alpha}_s^{{\overline {\rm MS}}}$, 
derived from various resummations of the BjPSR at
$Q^2_{\rm ph}\!=\!3 {\rm GeV}^2$.
Predictions corresponding
to $S_{\rm max}\!=\!0.241$ cannot be made with the TPS--PMS,
because the latter cannot be larger than
$(2/3)^{3/2} {\rho}_2^{-1/2}\!\approx\!0.233$, due to its
specific polynomial form.}
\label{tabl2}
\end{table}
Given are also predictions of such approximants when the
$\beta$--functions are TPS's ($c_k^{(j)}\!=\!0$ for $k \geq 4$). 
To highlight the importance of $c_3$--fixing, 
we included predictions of
these approximants (with ${\rm TPS}_{\beta}$)
when we set $c_3\!=\!0$ in them.
In addition, predictions of the following approximants
are included in Table \ref{tabl2}: 
TPS $S_{[2]}$ (\ref{TPSBj}) (NNLO TPS);
TPS $S_{[3]}$ with $r_3\!=\!128.05$ (${\rm N}^3{\rm LO}$ TPS);
off--diagonal Pad\'e approximants (PA's) $[1/2]_S$ and $[2/1]_S$;
square root of the diagonal PA (dPA) $[2/2]_{S^2}$,
which is based solely on the TPS $S_{[2]}$ (\ref{TPSBj}), 
as are the previous two off--diagonal PA's;
$[2/2]_S$ is the dPA constructed on the basis of
the ${\rm N}^3{\rm LO}$ TPS $S_{[3]}$ with $r_3\!=\!128.05$.
For $[2/2]_S$ and ${\rm N}^3{\rm LO}$ TPS we took the value 
$r_3\!=\!128.05$ (in ${\overline {\rm MS}}$, at RScl 
$Q_0^2\!=\!3 {\rm GeV}^2$) because then the $[1/2]$ PA
of the invariant Borel transform ${\overline B}_S$ (\ref{BT3})
predicts the correct ${\rm IR}_1$ pole 
$z_{\rm pole}\!=\!1/{\beta}_0$. 
Numbers in Table \ref{tabl2} are with four digits so that 
predictions of various methods can be easily compared.

Table \ref{tabl2} includes predictions for
$\alpha_s^{{\overline {\rm MS}}}(M_Z^2)$.
They were obtained from 
${\alpha}_s^{{\overline {\rm MS}}}(3 {\rm GeV}^2)$
by evolution via four--loop RGE, using the values of
the four--loop ${\overline {\rm MS}}$ coefficient 
$c_3(n_f)$ \cite{RVL}
and the corresponding three--loop matching
conditions \cite{Chetyrkinetal} for the flavor
thresholds. In the matching, we used the scale
$\mu(n_f)\!=\!\kappa m_q(n_f)$ 
above which $n_f$ flavors are active,
with $\kappa\!=\!2$, and $m_q(n_f)$ being the running 
quark mass $m_q(m_q)$ of the $n_f$'th flavor.
If increasing $\kappa$ from $1.5$ to $3$, 
the predictions for 
$\alpha_s^{{\overline {\rm MS}}}(M_Z^2)$ decrease by 
at most $0.15 \%$.
If we use $[2/3]_{\beta}$ instead of ${\rm TPS}_{\beta}$
in the evolution from $3 {\rm GeV}^2$ to $M_Z^2$,
${\alpha}^{{\overline {\rm MS}}}_s(M_Z^2)$ 
decreases by less than $0.04 \%$.

In Fig.~\ref{fig1Bjs} we present
various approximants as functions of 
${\alpha}^{{\overline {\rm MS}}}_s(M_Z^2)$.
Our, ECH and TPS--PMS approximants have TPS $\beta$--functions 
and $c_3\!=\!12.5, 17, 16$, respectively (by the described 
${\rm IR}_1$ pole requirement).
These three approximants, when the $\beta$--functions
are resummed by PA's, and $c_3\!=\!15.5, 20, 19$,
respectively (${\rm IR}_1$ pole),
are presented in Fig.~\ref{fig2Bjs}. 
The three approximants with 
${\rm TPS}_{\beta}$'s (from Fig.~\ref{fig1Bjs})
are included in Fig.~\ref{fig2Bjs} for comparison. 

If we reexpand the approximants in powers of $a_0$
(RScl $Q_0^2\!=\!Q^2_{\rm ph}$, in ${\overline {\rm MS}}$,
$n_f\!=\!3$), predictions for coefficient $r_3$
at $a_0^4$ of expansion (\ref{TPSBj}) are obtained. 
Our approximant $\sqrt{{\cal A}_{S^2}}(c_3)$,
with $c_3\!=\!15.5$, predicts 
$r_3\!=\!125.8\!-\!c_3^{{\overline {\rm MS}}}/
2\!+\!c_3 \!\approx\!130.8$,
and the ECH ${\cal A}_S^{{\rm (ECH)}}\!=\!a( \ln Q^2_{\rm ECH};
\rho_2, c_3,\ldots)$, 
with $c_3\!=\!20.$, predicts $r_3\!=\!129.9\!+\! 
(-\!c_3^{{\overline {\rm MS}}}\!+\!c_3)/2\!\approx\!129.4$.
Both predictions agree well with that
of \cite{KS} $r_3\!\approx\!129.9$ ($\approx\!130.$)
which was obtained from the ECH under the assumption 
$(-\!c_3^{{\overline {\rm MS}}}\!+\!c_3)\!\approx\!0$ 
(note that 
$c_3^{{\overline {\rm MS}}}\!\approx\!21.0$
was not known at the time \cite{KS} was written).

\section{Discussion and conclusions}

\begin{figure}[ht]
\centering\epsfig{file=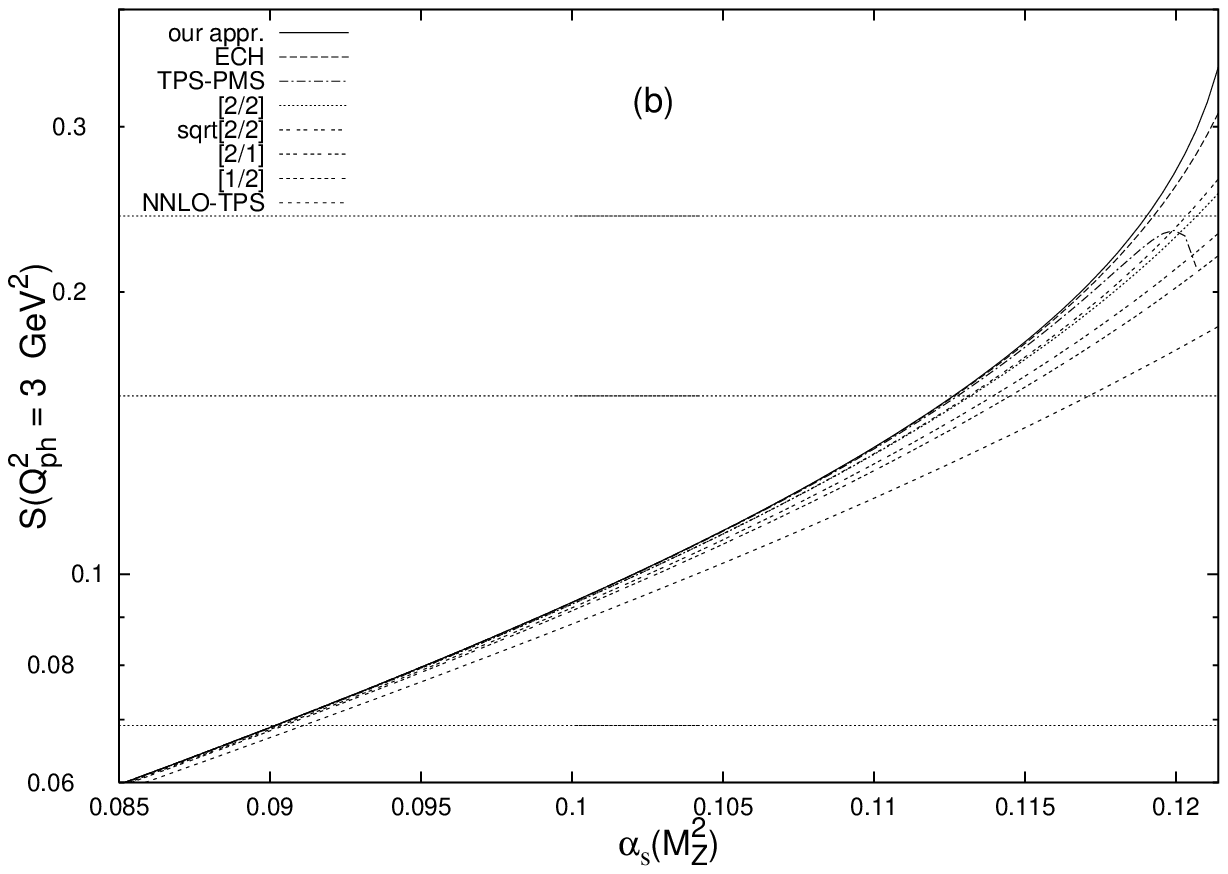,height=9.5cm}
\vspace{0.3cm}
\caption{Values of various approximants
(of Table \ref{tabl2}) as functions of 
${\alpha}^{{\overline {\rm MS}}}_s(M_Z^2)$.
The $\beta$--functions in our, ECH and TPS--PMS approximants 
have TPS form and the values of $c_3$ in them
were determined by the described ${\rm IR}_1$ pole requirement.
The experimental bounds (\ref{Bjdat}) 
$S_{\rm min}\!=\!0.069$, $S_{\rm max}\!=\!0.241$ and 
$S_{\rm mid}\!=\!0.155$ are indicated as horizontal lines.} 
\label{fig1Bjs}
\end{figure}

\begin{figure}[ht]
\centering\epsfig{file=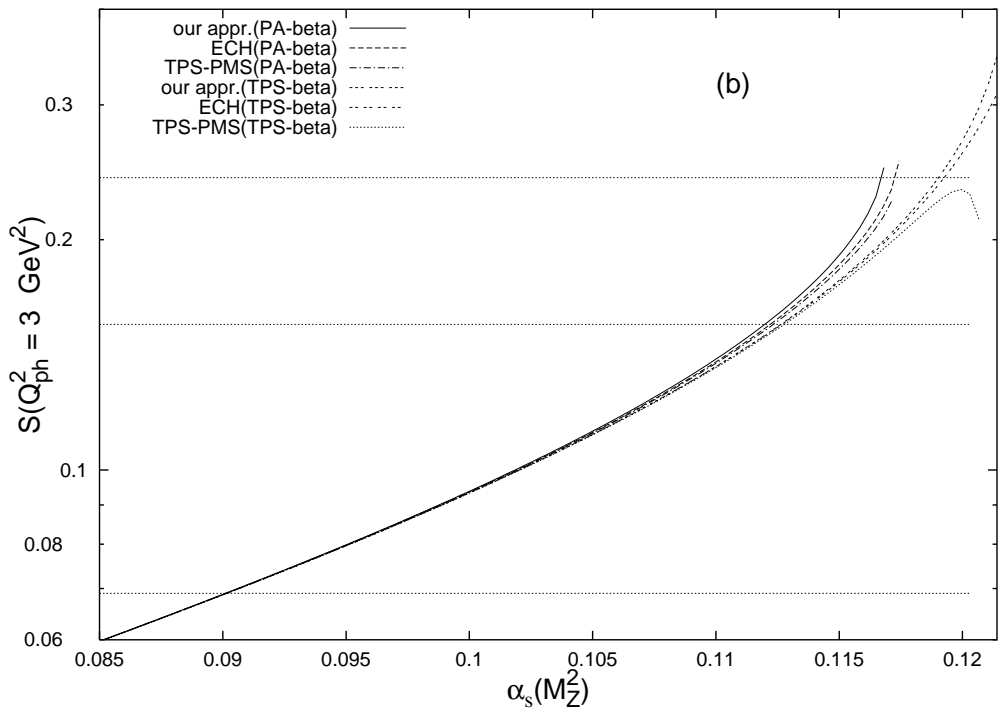,height=9.5cm}
\vspace{0.3cm}
\caption{\footnotesize 
As in Fig.~\ref{fig1Bjs}, but the $\beta$--functions
in our, ECH and TPS--PMS approximants are now resummed
by PA's and the values of $c_3$ subsequently determined 
by the ${\rm IR}_1$ pole requirement (see the text).
For comparison, the corresponding predictions from 
Fig.~\ref{fig1Bjs}, with ${\rm TPS}_{\beta}$'s, are included.}
\label{fig2Bjs}
\end{figure}

The results presented in Table \ref{tabl2} and in
Figs.~\ref{fig1Bjs}--\ref{fig2Bjs} show clearly that
nonperturbative effects, as reflected in the
mechanism of quasianalytic continuation
from the small--$a$ into large--$a$ regime
and in the presence of the leading infrared renormalon 
(${\rm IR}_1$) pole, play an important role in the
BjPSR at low photon transfer momenta
$Q_{\rm ph} \approx 1.73$ GeV. These effects
decrease the predicted value of 
${\alpha}_s^{\overline {\rm MS}}(M_Z^2)$
by very substantial amounts. 
Our approximant gives the BjPSR--prediction
${\alpha}_s^{\overline {\rm MS}}(M_Z^2) =
0.1120^{+0.0047}_{-0.0219}$ (see Table \ref{tabl2}).
Availability of additional data on polarized structure functions,
especially in the low--$x_{\rm Bj}$ regime,
may significantly reduce the uncertainties
of the BjPSR--predictions for
${\alpha}_s^{\overline {\rm MS}}(M_Z^2)$.

The present world average is
${\alpha}_s^{{\overline {\rm MS}}}(M_Z^2) =
0.1173 \pm 0.0020$ by Ref.~\cite{Hinchliffe},
and $0.1184 \pm 0.0031$ by Ref.~\cite{Bethke}.
The NNLO TPS predictions of the considered BjPSR 
($0.1171^{+0.0102}_{-0.0260}$, see Table \ref{tabl2})
cover the entire world average interval and more.
However, when the afore--mentioned two classes of
nonperturbative effects are taken into account,
e.g. via the use of our or the ECH approximants
and by the described $c_3$--fixing,
we obtain an upper bound 
${\alpha}_s^{{\overline {\rm MS}}}(M_Z^2)_{\rm max}
\approx 0.117$ -- see Table \ref{tabl2}.
But this upper bound does not surpass the central values of
the afore--mentioned world averages.
The central value of ${\alpha}_s^{{\overline {\rm MS}}}(M_Z^2)$
extracted from the BjPSR ($\approx\!0.112$)
is significantly lower than the world average.

What could be the reason for this?
One speculative possibility
would be that some of the Feynman diagrams
contributing to the ${\rm N}^3 {\rm LO}$ term
(yet unknown) of the BjPSR have a genuinely
new topology not appearing in the lower diagrams,
and that such new topology diagrams push the
predicted values of ${\alpha}_s^{{\overline {\rm MS}}}(M_Z^2)$
significantly upwards. The resummation methods based on the
NNLO TPS cannot ``foresee'' such contributions \cite{BLM,KS}.
In this context, we note that the afore--described
$c_3$--fixing in our, ECH and TPS--PMS approximants
enables these approximants to be based on more than just
the information contained in the NNLO TPS and in the RGE. 
However, since the location of the ${\rm IR}_1$ pole
can be determined by large--${\beta}_0$
considerations, the described $c_3$--fixing apparently
does not incorporate information from those possible 
higher--loop diagrams whose topologies are genuinely new.  

Another possible reason for the difference between
our ${\alpha}_s^{{\overline {\rm MS}}}$--predictions
and those of the world average could for example
lie in a hitherto underestimated relevance of
nonperturbative contributions and of higher order
perturbative terms in the numerical analyses of data
for some QCD observables. This possibility should be seen
also in view of the fact that (some) NNLO contributions
($\sim$$a^3$) are not yet theoretically known
for several of the quantities whose data
have been analyzed to predict the world average 
\cite{Hinchliffe,Bethke}.
However, lower values are allowed by
some recent analyses beyond the NLO:
${\alpha}_s^{\overline {\rm MS}}(M_Z^2) =
0.118 \pm 0.006$ \cite{Kataevetal}
from the CCFR data for $x_{\rm Bj} F_3$
structure function from ${\nu} N$ DIS (NNLO);
$0.112^{+0.009}_{-0.012}$ from
Gross--Llewellyn--Smith sum rule \cite{Bethke} (NNLO);
$0.115 \pm  0.004$ \cite{Hinchliffe} from lattice
computations.

From the theoretical point of view, we are
dealing with three types of resummation
approximants for NNLO TPS's of QCD observables
in the present paper:
\begin{enumerate}
\item
Pad\'e approximants (PA's)
provide an efficient mechanism of quasianalytic
continuation. However, they do not possess
RScl-- and RSch--invariance, although their
dependence on the RScl and on the leading
RSch--parameter $c_2$ is in general weaker than
that of the original TPS. In addition, 
they implicitly possess a $c_3$--dependence,
but this dependence has no special role since there is
also $c_2$-- and RScl--dependence.
\item
Grunberg's ECH and Stevenson's TPS--PMS methods
do not possess a strong mechanism of quasianalytic
continuation, except the one provided by the
RGE--evolution of the coupling parameter $a$
itself.\footnote{
In the one--loop limit, this
amounts to the $[1/1]$ PA quasianalytic
continuation for $a$ (ECH).}
This is so because these approximants do
not go beyond the polynomial form in terms
of the coupling parameter $a$. On the other hand,
these approximants do achieve RScl-- and 
$c_2$--independence, since they represent a judicious
choice of the RScl and of $c_2$ in the TPS. They 
possess a $c_3$--dependence.
\item
Our approximants provide an efficient mechanism
of quasianalytic continuation, since they reduce
to the diagonal PA expression $[2/2]_{S^2}^{1/2}$
in the one--loop limit 
(when all $c_k^{(0)}, c_k^{(j)}, c_k \mapsto 0$ for $k \geq 1$).
At the same time, they possess invariance under
the change of the RScl and of the leading RSch--parameter
$c_2$. They possess a $c_3$--dependence.
\item
The dependence on $c_3$ (and on $c_k$, $k \geq 4$)
parameters in our, ECH and TPS--PMS
approximants allows us to incorporate into them
important nonperturbative information about the
location of the leading IR renormalon pole.
Further, it allows us to use in these approximants
resummed $\beta$--functions (PA--type),
thus presumably additionally strengthening the
effects of quasianalytic continuation mechanism.
These approximants are then fully independent of the
RScl and RSch of the original TPS.
\end{enumerate}

The leading higher--twist term contribution to the BjPSR
($\sim$$1/Q^2_{\rm ph}$) \cite{htlit}--\cite{Braun},
or a part of it, is implicitly contained in
our approximant, as well as in the ECH and the
TPS--PMS, via the afore--mentioned $c_3$--fixing.
The described approach implicitly
gives an approximant--specific prescription 
for the elimination of the
(leading IR) renormalon ambiguity.
It is not clear which approximant accounts for the
$\sim$$1/Q^2_{\rm ph}$ terms in the best way.

A more detailed and extensive presentation of the subject
will appear shortly \cite{prep}.

\vspace{0.5cm}

\noindent{{\large \bf Acknowledgement:}}
\noindent
This work was supported
by the Korean Science and Engineering Foundation (KOSEF).

\noindent

\noindent

\end{document}